\documentclass{appolb}
\usepackage[dvips]{graphicx}

\begin{document}

\title{Kepler Map for H atom driven by microwaves with arbitrary polarization}
\author{Prot Pako\'nski \and Jakub Zakrzewski
\address{Instytut Fizyki im. Mariana Smoluchowskiego, Uniwersytet
	 Jagiello\'nski, ul.~Reymonta~4, 30--059~Krak\'ow, Poland}}
\maketitle
\markboth{P.~Pako\'nski, J.~Zakrzewski}
         {Kepler Map for H atom driven by microwaves\ldots}

\begin{abstract}
Dynamics of hydrogen atom driven by microwave field of arbitrary
polarization is approximated by the discrete mapping. The map
describes the change of dynamical variables from an aphelion
or a perihelion to the next one. The results are compared with
numerical simulation and previous approximations.
\end{abstract}

\section{Introduction}

The microwave ionization of Rydberg atoms, a standard topic in the
discussion of the classical--quantum correspondence
in classically chaotic systems \cite{Ca87,Ca88,Je91,Ko95},
brings still to us new challenges and surprises. The development
of experimental techniques allows one now to vary experimentally
the microwave polarization \cite{Fu90,Be97}. That in turn
enables an efficient control of the dynamics and is a challenging
problem for theoretical study \cite{Ri92,Ri97,Ok20,OD20}.
As it becomes now more and more clear, different quantum initial
states contribute predominantly to the onset of the ionization
when the microwave polarization changes \cite{Be97,Sa98}.

The full numerical quantal description of the problem becomes
within reach of the present days computers. Still simplified
models have contributed a lot to our understanding of the
mechanism of excitation of an atom  and its subsequent
ionization especially in term of classical chaos on one side and
dynamical localization in the quantum approach \cite{Ca88}.
Among such a simplified description a special place is
taken by the Kepler map \cite{Pe86,CG87,Go87,Ca88} where the
full continuous dynamics is approximated by a discrete
mapping. The name Kepler map was introduced in \cite{CG87}.
This map could be used both to classically estimate
the experimental ionization threshold as well as to give the
quantum mechanical predictions after an appropriate quantization
\cite{CG87,Ca88,Ka99,Be00}.

The main idea behind the mapping approach is an understanding that
the electronic motion on the Kepler orbit is the most sensitive
to an external microwave perturbation when the electron passes
in the vicinity of the nucleus. Such an	 approach was first
applied to a straight line orbit (one-dimensional approximation)
\cite{Pe86,CG87,Go87}. This approach was further developed
and quantized by Casati, Guarneri and Shepelyansky \cite{CG87,Ca88}.
The kicked-like dynamics assumes a Kepler electronic motion
between passages at perihelions and the kick every time
the electron is at perihelion. One may then obtain the map describing
the dynamics between the successive passages at perihelions, or,
as proposed in \cite{Ca88}, to make the map describing the change
of dynamical variables from an aphelion to the next one. To calculate
the kick strength the influence of microwaves over whole Kepler period
was integrated. This is why the Kepler Map may be seen also as the
approximation of the mapping generated by Poincar\'e section at radial
momentum $p=0$ \cite{Na90}. This argument shows that approximation
should be also valid when the electron orbit stops to be linear.
Further approximations \cite{Ca88} lead to a map equivalent to
a standard Chirikov map \cite{Li83}. The quantum mechanical
behavior of the latter is well understood \cite{Fi82} leading
to an explanation of classical--quantum differences in microwave
ionization of H atoms in terms of the Anderson (dynamical) localization.

The authors of \cite{Ca88} extended their analysis to the circular
microwave polarization. However, soon Nauenberg showed \cite{Na90}
that the dynamical variables in the derivation of the Kepler Map
are not canonically conjugate. He derived the Kepler Map
for H atom in circularly polarized microwaves.

We propose the construction of the Kepler Map for H atom subjected in
microwaves of arbitrary polarization. This construction allows us to
answer questions concerning ``canonicity'' of variables in previous works.
We analyze the limiting cases of the linear motion of the electron in
the same linear polarization of the microwave field, and the circular
polarization of the microwaves. Our new Kepler Map reduces to the
previous approaches in both cases. We compare the map with numerically
integrated equations of motion and study validity of the approximation.
Further, surprising limitations of such an approach are discussed.

\section{Construction -- one dimensional model}
The Hamiltonian of one dimensional (1D) hydrogen atom in microwave
fields reads
(in atomic units)
\begin{equation}
  H = \frac{p^2}{2}-\frac{1}{x}+Fx\cos\omega t
\end{equation}
where $(x,p)$ are the position and momentum variable, $F$ is the
amplitude of microwaves and $\omega$ their frequency. The motion
takes place for $x>0$. Perihelions of the motion described by this
Hamiltonian correspond to the point $x=0$, when the electron bounces
from the nucleus. Momentum $p=dx/dt$ then is infinite but the whole
Hamiltonian $H$ has finite limit for every $x \rightarrow 0$. To
eliminate the time dependence of $H$ we go to an extended phase space
\cite{Li83} with $(H,t)$ being the second pair of variables. The
new autonomous Hamiltonian is
\begin{equation}
  {\mathcal{H}} = \frac{p^2}{2}-\frac{1}{x}+Fx\cos\omega t-H = 0,
\end{equation}
leading to the following
equations of motion ($\xi$ is the new time of evolution)
\begin{equation}
  \begin{array}{ll}
    \frac{dx}{d\xi}=\frac{\partial{\mathcal{H}}}{\partial p}=p, &
    \qquad \frac{dp}{d\xi}=-\frac{\partial{\mathcal{H}}}{\partial x}
    =-\frac{1}{x^2}-F\cos\omega t , \\[4mm]
    \frac{dt}{d\xi}=\frac{\partial{\mathcal{H}}}{\partial H}=-1, &
    \qquad \frac{dH}{d\xi}=-\frac{\partial{\mathcal{H}}}{\partial t}
    =Fx\omega\sin\omega t .
  \end{array} \label{eqsm}
\end{equation}
We evaluate the change of $(H,t)$ variables from a perihelion
($x=0$) to the next one. These variables are canonically conjugate
so the resulting mapping should be a canonical
transformation. We introduce the eccentric anomaly $u$: $\cos u=1+2Ex$.
$E$ is the Kepler energy $E=\frac{p^2}{2}-\frac{1}{x}$ of the electron.
The parameter $u$ allows us to describe perihelions ($u=2k\pi$, $k$
integer) and aphelions ($u=\pi+2k\pi$) of the motion of the electron,
since the momentum is equal to $p=\sqrt{-2E\frac{1+\cos u}{1-\cos u}}$.
We can calculate $\frac{du}{d\xi}$
\begin{equation}
  \frac{du}{d\xi}= \frac{(-2E)^{3/2}}{1-\cos u}
    \left(1+\frac{F}{2E^2}(1-\cos u)\cos\omega t\right) \ .
\end{equation}
If $\frac{F}{E^2}<1$ the derivative $\frac{du}{d\xi}$ may be inverted and
the motion of the electron can be parametrized by  $u$
\begin{equation}
  \frac{dH}{du} = \frac{F\omega}{(-2E)^{5/2}}\,\frac{(1-\cos u)^2
    \sin\omega t}{1+\frac{F}{2E^2}(1-\cos u)\cos\omega t} \ ,
  \label{dHdu}
\end{equation}
\begin{equation}
  \frac{dt}{du} = \frac{-1}{(-2E)^{3/2}}\,\frac{1-\cos u}
    {1+\frac{F}{2E^2}(1-\cos u)\cos\omega t} \ . \label{dtdu}
\end{equation}
Integrating these equations to the first order in $F/E^2$ for
$0 \leq u \leq 2\pi$  we obtain 1D Perihelion Kepler Map
\begin{equation}
  H' = H + \frac{F\omega}{(-2H')^{5/2}}
    \int_0^{2\pi} (1-\cos u)^2\sin\omega\tau du ,
  \label{map1a}
\end{equation}
\begin{equation}
  t'\!=\!t - \frac{2\pi}{\omega_K}+\frac{F}{(-2H')^{7/2}}\!\int_0^{2\pi}\!
    \left[2(1-\cos u)^2-3\sin u(u-\sin u)\right]\!\cos\omega\tau du ,
  \label{map1b}
\end{equation}
where $\tau=(u-\sin u)/\omega_K+t$ is the time obtained from
equation~(\ref{dtdu}) integrated in zero order in $F/E^2$ with
$\tau=t$ for $u=0$ (at perihelion), $\omega_K=(-2H)^{3/2}$ is
the Kepler frequency. The condition of area preservation
leads to an implicit character of these equations (the right hand
side depends on $H'$). This map is generated by the function $G$
($H=\frac{\partial G}{\partial t}, t'=\frac{\partial G}{\partial H'}$)
\begin{equation}
  G(H',t) = H't+\frac{2\pi}{\sqrt{-2H'}}+\frac{F}{(-2H')^{5/2}}
    \int_0^{2\pi} (1-\cos u)^2\cos\omega\tau du . \label{gen1}
\end{equation}
The map~(\ref{map1a}),~(\ref{map1b})
coincides with Nauenberg 1D Kepler Map \cite{Na90}, although
his map is derived by evaluating the change of Kepler energy $E$
in place of $H$. For the 1D dynamics the perihelions always take
places at $x=0$, where $E=H$.

The alternative to the perihelion map is the aphelion map.
To obtain it we must integrate eqs. (\ref{dHdu}) and
(\ref{dtdu}) for $-\pi\leq u \leq\pi$ with the new function
$\tau=(u-\sin u+\pi)/\omega_K+t$, since now the condition is
$\tau=t$ for $u=-\pi$. The 1D Aphelion Kepler Map reads then as follow
\begin{eqnarray}
  H' & = & H + \frac{F\omega}{(-2H')^{5/2}}
    \int_{-\pi}^{\pi} (1-\cos u)^2\sin\omega\tau du , \label{map2a} \\
  t' & = & t - \frac{2\pi}{\omega_K}+\frac{F}{(-2H')^{7/2}}\int_{-\pi}^{\pi}
    (1-\cos u)^2 \nonumber \\
     &   & \left[5\cos\omega\tau-3\frac{\omega}{\omega_K}
    (u-\sin u+\pi)\sin\omega\tau\right] du .
  \label{map2b}
\end{eqnarray}
This map does not coincide with Nauenberg 1D Kepler Map \cite{Na90}
any more. Since these equations are related by the condition of area
preservation, we can concentrate on the first one (\ref{map2a}) only.
After integrating by part and some calculation we have
\begin{eqnarray}
  H' & = & H + \frac{F}{-2H'}\sin\left(\omega t+\pi\frac{\omega}
    {\omega_K}\right) \nonumber \\
     &   & \left[ 4\sin\pi\frac{\omega}{\omega_K}
    -\int_0^\pi\sin u\sin\left(\frac{\omega}{\omega_K}(u-\sin u)
    \right)du \right] . \label{map2c}
\end{eqnarray}
The remaining integral has nice asymptotic behavior for
$\omega\gg\omega_K$, this fact serves in \cite{Ca88} to
approximate Kepler Map by the standard-like mapping. In
the high frequencies regime we find
\begin{equation}
  H' = H + F\sin\left(\omega t+\pi\frac{\omega}{\omega_K}\right)
    \left[ \frac{4\sin\pi\frac{\omega}{\omega_K}}{-2H'}
    -\frac{2.58}{\omega^{2/3}} \right] , \label{map3a}
\end{equation}
\begin{equation}
  t' = t - \frac{2\pi}{\omega_K}-F \Bigl[ \frac{12\pi}{(-2H')^{7/2}}
    \cos\left(\omega t+2\pi\frac{\omega}{\omega_K}\right) +
\end{equation}
\begin{displaymath}
  + \frac{7.74\pi}{\omega^{2/3}(-2H')^{5/2}}
    \sin\left(\omega t+\pi\frac{\omega}{\omega_K}\right)\!+
    \frac{8}{\omega(-2H')^2}
    \cos\left(\omega t+\pi\frac{\omega}{\omega_K}\right)sin
		       \pi\frac{\omega}{\omega_K}
    \Bigr] . \label{map3b}
\end{displaymath}
As in standard map the dependence of $H'$ on $t$ is sinusoidal, but the
phase and the amplitude depend on the energy $H'$. The change of Kepler
energy $E$ from a aphelion to the next one is equal to
\begin{equation}
  E' = E - \frac{2.58F}{\omega^{2/3}}
    \sin\left(\omega t+\pi\frac{\omega}{\omega_K}\right), \label{map3c}
\end{equation}
but $t$ is not the canonical conjugate variable to $E$. Disregarding
this fact one could try to perform a canonical change of variables to
have new time $T=t+\pi/\omega_K$, keeping the energy $E$ fixed. Then
however, to express the mapping in these new variables, first it is
necessary to solve an implicit equation, which is an analog to eq.
(\ref{map3c}).

\section{Comparison of one dimensional dynamics}
\begin{figure}[hbt]
  \includegraphics[width=\textwidth]{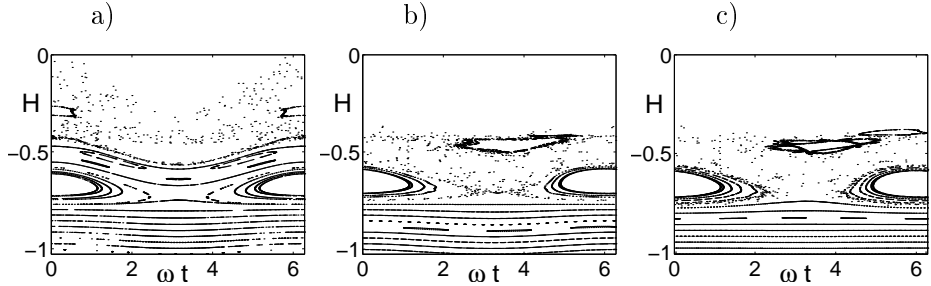}
  \caption{Comparison of numerically integrated dynamics of 1D
    hydrogen atom (left panel, denoted as a) with the Aphelion
    Kepler Map (middle panel, denoted as b) and the same map
    in the high frequency limit (right panel, c). The total
    energy $H$ is plotted versus $\omega t$ modulo $2\pi$. The
    amplitude of microwaves $F=0.02$ and frequency $\omega=1.6$.}
  \label{figh1}
\end{figure}

To verify our approximations  we have integrated numerically the
equations of motion of 1D model of hydrogen atom~(\ref{eqsm}) in
parabolic variables. Along each trajectory we have located all
aphelions or perihelions.
The figure~\ref{figh1} shows the total energies and times
($\omega t$ modulo $2\pi$) of subsequent passages by aphelion
generated from the numerical integration, 1D Aphelion Kepler
Map, and the same map in $\omega\gg\omega_K$ regime. To
find the next iteration of the map we solve numerically implicit
equations~(\ref{map2a}) and~(\ref{map3a}). For $H<-0.6$ the
trajectories, found by means of the map, follow ones coming from
numerical integration. The approximation breaks down for larger
$H$, because in derivation of the map we have approximated the
dynamics only to first order in $\frac{F}{H^2}$. When $H$
approaches to $0$ the implicit equation defining the map stops
to have unique solution, this fact explains the straight
horizontal lines for $H>-0.4$ on the maps picture.
Thus while exact dynamics may lead to an unbound diffusion
(and subsequently ionization), the map cannot be extended to that
regime.

\begin{figure}[hbt]
  \includegraphics[width=\textwidth]{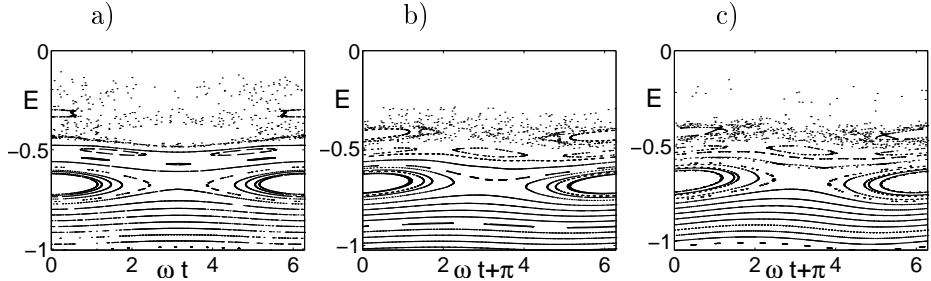}
  \caption{Comparison of numerically integrated dynamics of 1D
    hydrogen atom (left panel, denoted as a) with the Kepler Map
    of Casati et al. \protect\cite{Ca88} (middle panel, denoted
    as b) and the same map in high frequencies limit (right panel,
    c). The Kepler energy $E$ is plotted versus $\omega t$ modulo
    $2\pi$, the $\omega t$ variable of the map must be shifted by
    $\pi$ for correspondence with time. The amplitude of microwaves
    $F=0.02$ and frequency $\omega=1.6$.}
  \label{fige1}
\end{figure}

The similar comparison may be performed for the earlier
approximations of Kepler dynamics \cite{Pe86,Go87,Ca88}.
Figure~\ref{fige1} compares the aphelions found
from numerically integrated trajectories, the Kepler Map of
Casati et al. \cite{Ca88} and its high frequency limit version.
Here the Kepler energy $E$ is plotted versus $\omega t$ modulo
$2\pi$. The $\omega t$ variable for the maps are shifted by $\pi$
to restore the correspondence with the true time. The use of $E$
as the map variable and the shift of time comes from the definition
of the map in \cite{Ca88}. No problem with uniqueness of solutions
occurs when iterating these maps due to asymptotically ($\omega\gg
\omega_K$) explicit character of equations defining them.

\begin{figure}[hbt]
  \includegraphics[width=\textwidth]{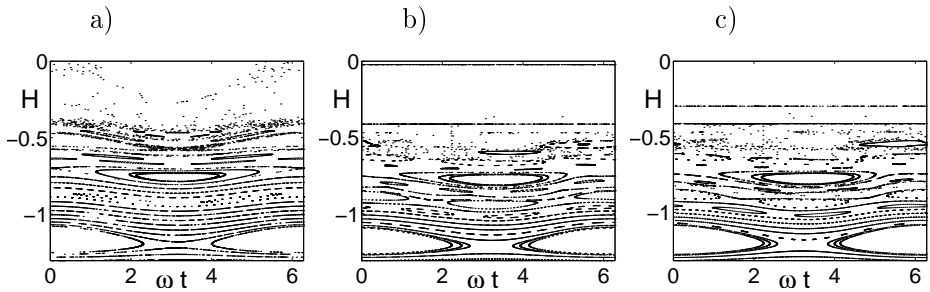}
  \caption{Same as Fig.~\protect\ref{figh1} but for $F=0.02$
    and $\omega=3.7$.}
  \label{figh2}
\end{figure}

\begin{figure}[hbt]
  \includegraphics[width=\textwidth]{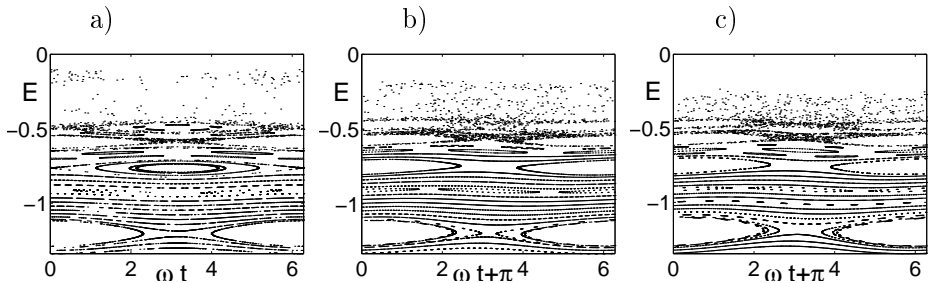}
  \caption{Same as Fig.~\protect\ref{fige1} but for $F=0.02$
    and $\omega=3.7$.}
  \label{fige2}
\end{figure}

A similar picture  comparing the 1D dynamics with the Kepler
Map is shown also  for higher frequency $\omega=3.7$. In
figure~\ref{figh2} the trajectories generated from the maps
follow the numerically integrated ones for small $H$. The
higher resonances 3:2 and 4:3 ($\omega:\omega_K$) appears
stronger and in a different position in the discrete dynamics
of the map than in the continuous case. Figure~\ref{fige2}
shows that for same frequency $\omega=3.7$ the canonical
transformation of non canonical variable $(E,t)$ may misplace
some characteristic part of the phase portrait. The resonance
2:1 is moved by $\pi$ in the $\omega t+\pi$ variable (the $\pi$
in the variable comes from the definition of the map).

\begin{figure}[hbt]
  \centering \mbox{\includegraphics[width=0.7\textwidth]{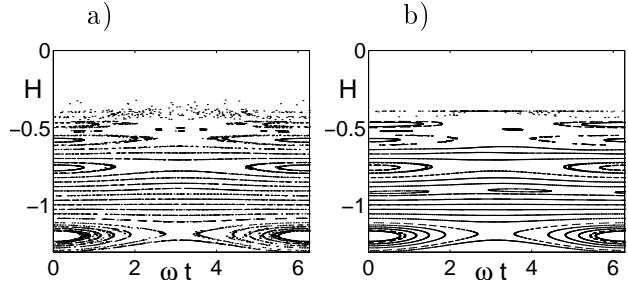}}
  \caption{Comparison of numerically integrated dynamics of 1D
    hydrogen atom (left panel) with the Perihelion Kepler Map
    (right). The total energy $H$ is plotted versus $\omega t$
    modulo $2\pi$. The amplitude of microwaves $F=0.02$ and
    frequency $\omega=3.7$.}
  \label{figh3}
\end{figure}

We have observed that Perihelion Kepler Map approximates the
numerical dynamics of the model for a somewhat larger range
of the energies $H$. The figure~\ref{figh3} shows the Poincar\'e
sections generated from perihelions found from numerically
integrated trajectories and found by means of the Kepler Map
(as given by the eqs.~(\ref{map1a}), (\ref{map1b})). The better
approximation for larger total energy $H$ may come from the fact,
that at perihelions the influence of perturbation is maximal,
the variables change rapidly. The map, obtained by the integration
of the influence of the microwaves over the Kepler period, should
be better if the interval of integration do not cross the
perihelion. Still the map suffers from the same drawback ---
it is valid locally only, given by implicit equations, which
does not have to (and typically do not) yield a unique solution
far from the initial conditions. Thus no prediction e.g. for the
ionization threshold may be made basing on the ``proper'',
canonical Kepler map.

\section{Construction -- two dimensional model}
The two dimensional model of hydrogen atom may be perturbed by the
microwave field of any polarization. The Hamiltonian of the autonomous
system in the extended	phase space reads
\begin{equation}
  {\mathcal{H}} = \frac{p^2}{2}+\frac{l^2}{2r^2}-\frac{1}{r}+
    Fr(\cos\phi\cos\omega t+\alpha\sin\phi\sin\omega t)-H = 0.
\end{equation}
We use radial coordinate $r$, $\phi$ and $t$ as the position variables and
$p$, $l$, $H$ as the corresponding momenta. The parameter $\alpha$ describes
polarization of the microwaves, it ranges from 0 (linear polarization) to
1 (circular one). In analogy to 1D model we introduce the eccentric anomaly
defined by $e\cos u=1+2Er$, where $E$ is the Kepler energy $E=\frac{p^2}{2}
+\frac{l^2}{2r^2}-\frac{1}{r}$ of the atom and $e$ is the eccentricity
$e=\sqrt{1+2El^2}$. The points of Poincar\'e section at $p=0$ (perihelions
and aphelions) are described by $u=k\pi$ with $k$ integer, since
\begin{equation}
  p = \sqrt{-2E}\frac{e\sin u}{1-e\cos u} \ .
\end{equation}
If the perturbation is small $F/E^2 \ll 1$, we can find $\frac{dH}{du}$ and
$\frac{dl}{du}$ and integrate them for $0 \leq u \leq 2\pi$ to first order
in $F/E^2$
\begin{equation}
  H' = H + \frac{F\omega}{(-2H')^{5/2}} \int_0^{2\pi} (1-e'\cos u)^2
    (\cos\theta\sin\omega\tau-\alpha\sin\theta\cos\omega\tau) du ,
  \label{map4a}
\end{equation}
\begin{equation}
  l' = l + \frac{F}{(-2H')^{5/2}} \int_0^{2\pi} (1-e'\cos u)^2
    (\sin\theta\cos\omega\tau-\alpha\cos\theta\sin\omega\tau) du ,
  \label{map4b}
\end{equation}
where functions $\tau=(u-e'\sin u)/\omega_K+t$, $\omega_K=(-2H')^{3/2}$
and $\theta=\chi(u)+\phi$,
\begin{equation}
  \chi(u)=\int_0^u\,\frac{\sqrt{1-e'^2}}{1-e'\cos u'}du' \ , \quad
  \sin\chi(u)=\frac{\sqrt{1-e'^2}\sin u}{1-e'\cos u} \ ,
\end{equation}
the eccentricity $e'=\sqrt{1+2H'l'^2}$. The equations~(\ref{map4a})
and~(\ref{map4b}) are generated by the same function
\[ G(H',l',t,\phi) = H't+l'\phi+\frac{2\pi}{\sqrt{-2H'}}+2\pi l'+ \]
\begin{equation}
  + \frac{F}{(-2H')^{5/2}} \int_0^{2\pi} (1-e'\cos u)^2
    (\cos\theta\cos\omega\tau+\alpha\sin\theta\sin\omega\tau) du .
  \label{gen4}
\end{equation}
This generating function defines a 2D Perihelion Kepler Map for hydrogen
atom perturbed by microwaves of arbitrary polarization\footnote{The very
same procedure is possible in 3D where $l^2$ should be replace by the
total angular momentum squared $L^2$.}.

Two limiting cases of the transformation generated by~(\ref{gen4})
may be verified. The linear motion of the electron along the axis
of linearly polarized  microwave field corresponds to $l=0$,
$\sin\phi=0$ and $\alpha=0$. One can check that the implicit
equation~(\ref{map4b}) for $l'$ has the solution $l'=l=0$, the
$\sin\phi'$ remains 0, because $\phi$ changes by $2\pi$, and the
generating function~(\ref{gen4}) reduces to the 1D Perihelion
Kepler Map~(\ref{gen1}).

For circularly polarized microwaves $\alpha=1$ and the
integrated function in~(\ref{gen4}) reduces to $(1-e'\cos u)^2
\cos(\theta-\omega\tau)$. It depends on $t$ and $\phi$ via the
common variable $\phi-\omega t$. We may canonically change the
variables to have $\tilde\phi=\phi-\omega t$ and $\tilde t=t$,
so $\tilde H=H+\omega l$ and $\tilde l=l$. The generating
function of new variables reads as follows
\begin{equation}
  G(\tilde H',\tilde l',\tilde t,\tilde\phi) = \tilde H'\tilde t+
    \tilde l'\tilde \phi+\frac{2\pi}{\sqrt{-2(\tilde H'-\omega\tilde l')}}
    +2\pi\tilde l'+ 
\end{equation}
\begin{displaymath}
  \frac{F}{(-2(\tilde H'-\omega\tilde l'))^{5/2}} \int_0^{2\pi}
    (1-e'\cos u)^2\cos\left(\chi(u)-\frac{\omega}{\omega_K}(u-e\sin u)
    +\tilde\phi\right) du . \label{gen4a}
\end{displaymath}
$\tilde H$ is invariant under the transformation~(\ref{gen4a}). It is
the autonomous Hamiltonian of the system in the coordinate frame
rotating with the microwave frequency. The function~(\ref{gen4a})
generates in variables $(\tilde l,\tilde\phi)$ Nauenberg Kepler Map for
circularly polarized microwaves \cite{Na90}.

\section{Comparison of two dimensional dynamics}
\begin{figure}[hbt]
  \includegraphics[width=\textwidth]{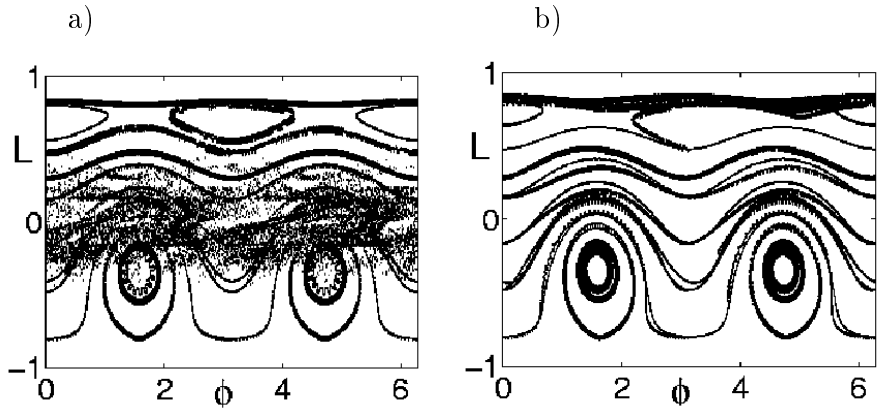}
  \caption{Comparison of numerically integrated dynamics of 2D
    hydrogen atom (left panel a) with Perihelion Kepler Map (right
    panel b). The angular momentum $L$ is plotted versus the radial
    angle $\phi$, this is the projection of Poincar\'e surface of
    section so the trajectories may cross. The amplitude of microwaves
    $F=0.01$, frequency $\omega=2$ and the polarization $\alpha=0.5$.}
  \label{figl1}
\end{figure}

\begin{figure}[hbt]
  \centering \mbox{\includegraphics[width=0.7\textwidth]{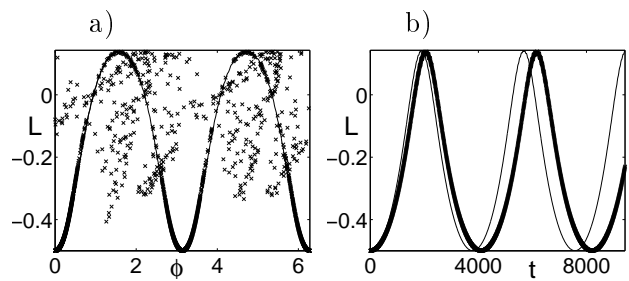}}
  \caption{The dependence of angular momentum $L$ on radial
    angle $\phi$ (a) and time $t$ (b) in perihelions for a generic
    trajectory. The crosses denote $\phi$ values found from the
    numerical integration of equations of motion for 2D model of
    H atom and the solid line is the result of applying 2D Kepler
    Map. For $L \approx 0$ the radial angle at perihelion is not
    well defined, this results in isolated points in the left
    plot. The $t$ dependence does not show a similar problem. The
    amplitude of microwaves is	$F=0.01$, frequency $\omega=2$
    and the polarization $\alpha=0.5$.}
  \label{figl2}
\end{figure}

The Poincar\'e sections generated by 2D Perihelion Kepler Map are
four dimensional. The fast dynamics in $(H,t)$ variables
is accompanied by the secular motion in $(L,\phi)$.
Figure~\ref{figl1} shows the $(L,\phi)$ projection of Poincar\'e
section generated from perihelions found by numerical integration
of equations of motion for 2D model of hydrogen atom  and found
from 2D Perihelion Kepler Map.
All trajectories start with $H=-0.5$ and $t=0$.
Small oscillations in trajectories come from the rapid motion
in $(H,t)$ variables.
The shadow of isolated points placed near $L=0$ on the Poincar\'e
section with numerical trajectories is caused by fact, that the
radial angle, at which the perihelion occurs, stops to be well
defined for very elongated orbits. The map does not suffer from this
problem. To demonstrate this feature we plot (Figure~\ref{figl2})
the dependence $L(\phi)$ and $L(t)$ in perihelions for a generic
trajectory crossing $L=0$ with same parameters as
in Figure~\ref{figl1}. The time dependence of angular momentum
shows that the Kepler Map predicts good values of $L$ at
subsequent perihelions. The small error in location of
perihelions in the numerically integrated trajectory results
in a much bigger error of radial angle, due to a very rapid
motion of electron near perihelion, when the orbit is
almost linear. The time dependence of the angular momentum $L$
demonstrates that the Kepler Map evolves a little faster
than the true dynamics.

\section{Conclusions}
The construction of approximate dynamics of hydrogen atom
driven by microwave field of arbitrary polarization by means
of a canonical mapping has been presented. The approximation
-- Kepler Map was compared with numerically integrated dynamics
of perturbed H atom and with previously constructed mappings
\cite{Pe86,Go87,Ca88}. Although the application of the map
requires solving an implicit integral equation, which is not
faster than integration of equations of motion, the map may be
used to reduce the dimensionality of dynamics and to an
approximate quantization.

On the other hand the maps obtained suffer from a major
drawback. While being accurate in the vicinity of initial
conditions they cease to be so when the electron energy
increases significantly due to the diffusion. The
effective microwave amplitude becomes then quite strong
and the maps, obtained in the first order in $F/E^2$, no
longer approximate the dynamics. Even worse, they do not yield
a unique solutions in that regime. This behavior may be
contrasted with the original Kepler map due to Casati
and coworkers \cite{Ca88} which while not canonical and,
strictly speaking, incorrect yields some estimate for
the onset of chaotic diffusion and ionization. Its other
great advantage is the simplicity of its form. Still, it is
our opinion, that one should be very cautious while trying to
extract any quantitative information from this celebrated mapping
due to the shown inconsistencies in its construction.

The support by the Polish Komitet Bada\'n Naukowych under
the grant No. 2~P03B~009~18 (P.P.) and 2~P03B~009~15 (J.Z.) is
acknowledged.

\end{document}